\documentclass[11pt]{article} 

\usepackage[utf8]{inputenc} 

\usepackage{geometry} 
\usepackage{graphicx} 

\usepackage{booktabs} 
\usepackage{array} 
\usepackage{paralist} 
\usepackage{verbatim} 
\usepackage{subfig} 

\usepackage{fancyhdr} 
\usepackage{sectsty}
\allsectionsfont{\sffamily\mdseries\upshape} 

\usepackage[nottoc,notlof,notlot]{tocbibind} 
\usepackage[titles,subfigure]{tocloft} 
\usepackage{amsfonts}
\usepackage{amsmath}
\usepackage{geometry}

\title{Brief Article}
\author{The Author}

\begin{document}
\title{Variable Tension, Large Deflection Ideal String Model For Transverse
Motions}
\author{Namik Ciblak \\
Department of Mechanical Engineering\\
Yeditepe University, Istanbul, Turkey\\
E-Mail: nciblak@yeditepe.edu.tr\\
http://me.yeditepe.edu.tr}
\maketitle

\begin{abstract}
In this study a new approach to the problem of transverse vibrations of an
ideal string is presented. Unlike previous studies, assumptions such as
constant tension, inextensibility, constant crosssectional area, small
deformations and slopes are all removed. The main result is that, despite
such relaxations in the model, not only does the final equation remain
linear, but, it is exactly the same equation obtained in classical
treatments. First, an "infinitesimals" based analysis, similar to historical
methods, is given. However, an alternative and much stronger approach,
solely based on finite quantities, is also presented. Furthermore, it is
shown that the same result can also be obtained by Lagrangian mechanics,
which indicates the compatibility of the original method with those based on
energy and variational principles. Another interesting result is the
relation between the force distribution and string displacement in static
cases, which states that the force distribution per length is proportional
to the second spatial derivative of the displacement. Finally, an equation
of motion pertaining to variable initial density and area is presented.

\end{abstract}

\section{Introduction}

The well known ideal string model used for analysis of transverse string
vibrations has been around for almost three centuries. This historical
approach is also used as one of the first examples in elementary or advanced
texts on partial differential equations (PDE) and mathematical physics. The
resulting PDE is usually solved by Bernoulli's separation method which
yields two second order linear ordinary differential equations, one for
spatial dimension and one for temporal. For certain boundary conditions the
total solution can be constituted in form of Fourier series. The uniqueness
of the solution is also proven without much difficulty \cite{Sagan}. This
famous wave equation is%
\begin{equation}
y_{xx}=\frac{1}{c^{2}}y_{tt}
\end{equation}%
where subscripts denote partial differentiation.

Aside from being one of the simplest and exemplary PDEs, the importance of
the ideal string equation also stems from a few other reasons. First, the
analytical solutions seem to agree with experimental results with surprising
accuracy. As definitive examples, one may recall the case of temporal
natural frequencies for a finite string. The predicted principal natural
frequencies in such cases are confirmed with extremely high accuracy in, for
example, stringed musical instruments \cite{Benson}. However, there is a
second reason that contributes much more to the importance of the ideal
string model as perceived in physics, mathematics, and engineering
community: the fact that it led to a new, general, and very rich subject
matter covering interesting topics in wave mechanics from non-dispersive
waves, standing waves, to wavelets; eventually to string theories in physics.

Generality of the theory was especially remarked by the fact that, instead
of solutions in series form, any general solution to the ideal string
problem can be represented as $f(x+ct)+g(x-ct)$ (D'Alembert's solution)
which is the sum of two non-dispersive "waves" traveling in opposite
directions with a constant speed of $c$ (see, for example, \cite{Sagan}, 
\cite{Benson}).

Nevertheless, what is probably the most surprising point is that this
accurate and versatile model is obtained only after \textit{quite strict}
assumptions and approximations -- physical as well as mathematical. At
first, these assumptions may seem reasonable. The reason is probably
psychological, which forces us to believe that such a beautiful and useful
result that stood the scrutiny of so many authorities, for more than two
hundreds of years, cannot be too far from being correct. Yet, in this study
we show that most of the assumptions and approximations used in arriving at
the classical string model are either unnecessary or contradictory, or both.
What is more, even in the absence of such assumptions and approximations one
gets, in a surprisingly straightforward manner, the very same PDE as that of
the classical model.

The new model proposed here also has its own limitations, too. These are
discussed in detail in final sections and conclusions are drawn. There are
still open problems and avenues for further developments.

\section{Assumptions and Approximations of the Classical Model}

The assumptions and approximations utilized in developing the classical
ideal string model have either physical or mathematical nature, or both.
These are presented and discussed below.

\begin{enumerate}
\item \textbf{Perfect Flexibility}. This is a physical model assumption. In
this model, string is assumed not to resist any bending moments. In other
words, the string is able to bend to any angle, at any point, without
creating any internal resistance. This means that the string only
experiences internal tension along its length. There are studies which
include the bending resistance effect, especially in analyses of musical
instruments. However, they lead to complicated forth order PDEs, albeit
closer to the reality (see \cite{Gottlieb}, \cite{Lai}, and \cite{Rao}). As
this is not the aim of this study, we shall also adopt this assumption.

\item \textbf{Constancy of Density}. This is an understandable and
acceptable approximation to the real strings, made of nearly homogeneous
materials, manufactured under controlled and consistent conditions. However,
as far as a local analysis is considered, this is really not necessary to
obtain a workable equation of motion. In this study, we, at first, allow
this restriction for simplicity, then present a variable density case.

\item \textbf{Initial Uniformity of Cross-sectional Area}. This pertains to
the condition of the string at rest. As in the case of density, this will be
shown to be an unnecessary physical approximation. However, more
importantly, this is certainly not the case in reality. In manufacturing of
real strings the cross-sectional area of the string, in both shape and
dimensional characteristics, will at best be a stochastic quantity dependent
on position. We also relax this condition in this study. Note that, in
almost all treatments, this assumption is never stated explicitly. Instead,
it is blended into the constancy of the density by way of considering the
linear density, rather than the volumetric density, the former being the
product of the latter and the area. However, in order to demonstrate the
main result of this study, \emph{at first, this assumption is allowed}.

\item \textbf{Constancy of Cross-sectional Area}. This is about the
condition of cross-sectional area of the string in motion. Classical
treatments, either explicitly or tacitly, assume that the area remains the
same throughout the whole motion. We shall show that this assumption is
impossible to retain if the following assumptions are to be revoked.

\item \textbf{Constancy of Tension}. This is both an assumption and an
approximation -- unsustainable in either case. First, anyone who has ever
played a stringed musical instrument will tell you that this is not the
case. As one plucks the string of a guitar, an undeniable increase in the
tension is felt -- up to the point of fracturing the wire. Thus, even in the
initial condition, the tension will be different from that at rest. This
phenomenon manifests itself as a changing pitch of the sound from the first
plucked instant to the end -- and musicians sometimes use this fact to
create beautiful artistic effects. Hence, as an approximation this is not
really acceptable. However, as the aim of this study is not to develop a
string model closer to the reality, this first objection is really not the
dominating one. Rather, there is a second reason: that not only is this
restriction mathematically unnecessary, but it is also inconsistent. In this
study, we let the tension vary along the length of the string and show that
we still obtain the classical equation of motion.

\item \textbf{Constancy of Length (Inextensibility)}. The case for this is
very much similar to the constancy of the tension. If a string is to be
given an initial displacement, its length cannot remain the same. One may
argue that this is only an approximation acknowledging the fact that the
displaced length is very close to the length at rest. We shall show that
this is an unnecessary mathematical approximation. Furthermore, we show that
if the assumption of constant tension is to be revoked than the constancy of
length cannot be retained, and vice versa.

In some other treatments, this assumption is not used. Instead, properly,
the string is taken to be perfectly elastic (see \cite{Weinstock}). Then,
they naturally get the same equation of motion. Nevertheless, they still
retain the other unnecessary assumptions such as the constancy of tension,
small displacements and slopes, and so on.

\item \textbf{Small Displacements and Slopes}. These are derived based on
the fourth, fifth, and the sixth cases above. Sometimes these are also used
as justifications of the formers.
\end{enumerate}

There are studies involving significantly larger displacements with variable
string length. However, they all are based on the constancy of tension and
lead to non-linear PDEs (\cite{Gottlieb}, \cite{Lai}, \cite{Rao}). The
results of this study show that such studies are inconsistent. We shall show
that assumptions fourth through sixth are equivalent. That is, for example,
one cannot argue that the length can vary while the tension is kept
constant, and vice versa. Also, removing fourth to sixth also undoes the
seventh.

\section{Variable Tension and Length Model}

\begin{figure}[htb!]
    \centering
      \includegraphics[width=0.65\textwidth]{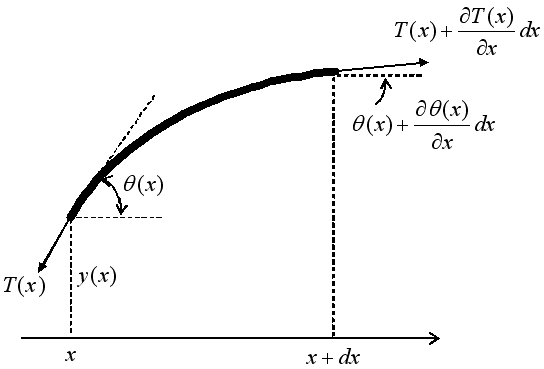}
    \caption{An infinitesimal string piece in tranverse motion.}
  \label{Fig1}
\end{figure}

A portion of a string in transverse motion is shown in Figure \ref{Fig1}.
The rest length of the element is $dx$. The variable internal tension is
shown at cut locations, with a first order variation. For now, we shall
assume constant density. Applying Newton's Second Law to this piece, in both 
$x$ and $y$ directions, one gets the following equations of motion.%
\begin{eqnarray}
\left( T+dT\right) \cos (\theta +d\theta )-T\cos (\theta ) &=&0 \\
\left( T+dT\right) \sin (\theta +d\theta )-T\sin (\theta ) &=&a\ dm
\end{eqnarray}%
where $dm$ is the total mass of the piece and $a$ is its acceleration of its
center of mass (rotational effects are ignored). If a first order
approximation is performed on%
\begin{equation}
\cos (\theta +d\theta )=\cos (\theta )\cos (d\theta )-\sin (\theta )\sin
(d\theta )
\end{equation}%
based on $\cos (d\theta )\approx 1$ and $\sin (d\theta )\approx d\theta $,
one may argue that%
\begin{equation}
\cos (\theta +d\theta )\approx \cos (\theta )-\sin (\theta )d\theta 
\end{equation}%
which leads to%
\begin{equation}
\cos \theta dT-T\sin \theta d\theta =0
\end{equation}%
\begin{equation}
\frac{dT}{T}=\tan \theta d\theta 
\end{equation}%
in which second order differentials are neglected. Now, the last line can be
easily integrated to give%
\begin{equation}
T=C\left\vert \sec \theta \right\vert ,\qquad (C>0)
\end{equation}%
which is the solution of the tension in terms of the slope angle. This is
the part that is overlooked in previous studies. Without this form, a
variable length assumption gives rise to non-linearity. It is this
not-so-nice-looking result which actually restores the linearity of the
final result.

Note that the absolute value operation automatically ensures the
non-negativeness of the tension. However, after a closer examination, it can
be shown to be unnecessary. When a section of $y(x,t)$ is obtained by left
and right cuts, the angles of tangents at ends can also be viewed as the
angle describing the angle of the tension vector. If one simply recalls that 
$y(x,t)$ is tacitly assumed to be a single valued function, the tension
vector at left cut may only point towards the second and third quadrants,
and the one at right end towards first and forth quadrants. Based on the way
we defined the angles at each end, and the fact that $y$ is never allowed to
be multi-valued, one concludes that $-\pi /2\leq \theta \leq \pi /2$, at any
cut. Thus, $\cos \theta \geq 0$ and%
\begin{equation}
T=C\sec \theta   \label{tension}
\end{equation}

Now, we concentrate on the second equation. Again, after expanding the
trigonometric terms and applying first order approximations, one gets%
\begin{eqnarray}
\left( T+dT\right) \left( \cos d\theta \sin \theta +\sin d\theta \cos \theta
\right) -T\sin (\theta ) &=&a\ dm \\
\frac{dT}{d\theta }\sin \theta +T\cos \theta  &=&a\frac{dm}{d\theta }
\end{eqnarray}%
Using the solution for the tension:%
\begin{eqnarray}
C\tan ^{2}\theta +C &=&a\frac{dm}{d\theta } \\
C\sec ^{2}\theta  &=&a\frac{dm}{d\theta }
\end{eqnarray}%
is obtained. Next, we invoke the condition that all points of the string
move in transverse direction. This simply implies that no mass is moving in
or out of the region considered -- \emph{a case of conservation of mass}.
Then,%
\begin{equation}
dm=\rho A(x)ds=\rho A_{0}dx
\end{equation}%
where $\rho $ is the density, $A(x)$ is the cross-sectional area in current
condition, $A_{0}$ is the uniform cross-sectional area at rest, and $ds=%
\sqrt{1+\left( \frac{\partial y}{\partial x}\right) ^{2}}dx$ is the current
length of the piece. Also, the acceleration $a$ can be approximated by $%
\frac{\partial ^{2}y}{\partial t^{2}}$, since taking $\frac{\partial ^{2}y}{%
\partial t^{2}}+\frac{\partial }{\partial x}\left( \frac{\partial ^{2}y}{%
\partial t^{2}}\right) dx$, or any average of these, would lead to the same
in the limit, assuming the smoothness of $\frac{\partial ^{2}y}{\partial
t^{2}}(x)$, of course. With these, the final equation becomes%
\begin{equation}
\frac{d\theta }{dx}\sec ^{2}\theta =\frac{\rho A_{0}}{C}\frac{\partial ^{2}y%
}{\partial t^{2}}
\end{equation}%
Now, using the fact that $\frac{\partial y}{\partial x}=\tan \theta $,
keeping time fixed, and using chain rule, one has%
\begin{equation}
\frac{\partial ^{2}y}{\partial x^{2}}=\frac{d}{dx}\left( \frac{dy}{dx}%
\right) =\frac{d}{d\theta }\left( \tan \theta \right) \frac{d\theta }{dx}=%
\frac{d\theta }{dx}\sec ^{2}\theta   \label{yxxsectheta}
\end{equation}%
Therefore, the result becomes%
\begin{eqnarray}
\frac{\partial ^{2}y}{\partial x^{2}} &=&\frac{\rho A_{0}}{C}\frac{\partial
^{2}y}{\partial t^{2}} \\
y_{xx} &=&\frac{1}{c^{2}}y_{tt}
\end{eqnarray}%
which is the well known classical linear wave equation, apart from the fact
that nothing has been said about the nature of the constant $C$. This is the
main result of this study.

\subsection{An Alternative Method Leads To Much Stronger Result}

The main result can actually be obtained in a nicer way: without arguing
limits, infinitesimals, or first order approximations. We simply start with
a \emph{really} finite string piece. Let $T_{1}$ and $T_{2}$ be the
tensions, and, $\theta _{1}$ and $\theta _{2}$ be the angles at left and
right cuts, respectively. Then the force balance in $x$ direction dictates%
\begin{equation}
T_{1}\cos \theta _{1}=T_{2}\cos \theta _{2}
\end{equation}%
This relation must hold for any segment (hence for any pair of end points $%
x_{i},x_{j}$), at all times. Letting $f(x,t)=T(x,t)\cos (\theta (x,t))$, the
fact that $f(x_{i},t)=$ $f(x_{j},t)$ for all $x_{i},x_{j}$ leads to $%
f(x,t)=g(t)$, and in turn, the fact that $g(t_{i})=$ $g(t_{j})$ for all $%
t_{i},t_{j}\geq 0$ leads to the conclusion that $g(t)$ must be a constant.
Hence, the result is%
\begin{equation}
T\cos \theta =C
\end{equation}%
which, in a much more direct manner, gives the previous solution for the
tension.

In obtaining this result we have not employed any differentials,
infinitesimal elements, or approximations. Therefore, simply in order to be
compatible with Newton's Second Law, regardless of such details as whether
bending is included or not, a constitutive model is utilized or not, and so
on, and regardless of how complicated or higher order the model is, \textit{%
any string model must conform to this result}, provided that only transverse
motions are allowed. Of course, we are assuming that there are no shear
forces at cut ends.

A similar approach can be applied for transverse motions. The force balance
for the finite string piece in transverse direction gives%
\begin{equation}
T_{2}\sin \theta _{2}-T_{1}\sin \theta
_{1}=\int\limits_{x_{1}}^{x_{2}}y_{tt}dm
\end{equation}%
Now, we use the identity%
\begin{equation}
\int\limits_{x_{1}}^{x_{2}}T\cos \theta d\theta =\left. T\sin \theta
\right\vert _{x_{1}}^{x_{2}}-\int\limits_{x_{1}}^{x_{2}}\frac{dT}{d\theta }%
\sin \theta d\theta
\end{equation}%
which gives%
\begin{equation}
T_{2}\sin \theta _{2}-T_{1}\sin \theta
_{1}=\int\limits_{x_{1}}^{x_{2}}\left( T\cos \theta +\frac{dT}{d\theta }%
\sin \theta \right) d\theta
\end{equation}%
Here, we did not use partial derivatives since these relations must hold at
any given time. Using the above identity one gets,%
\begin{equation}
\int\limits_{x_{1}}^{x_{2}}y_{tt}dm=\int\limits_{x_{1}}^{x_{2}}\left(
T\cos \theta +\frac{dT}{d\theta }\sin \theta \right) d\theta
\end{equation}%
Now, by applying $T\cos \theta =C$,%
\begin{equation}
\int\limits_{x_{1}}^{x_{2}}y_{tt}dm=\int\limits_{x_{1}}^{x_{2}}\left(
C+C\tan ^{2}\theta \right) d\theta =\int\limits_{x_{1}}^{x_{2}}C\sec
^{2}\theta d\theta
\end{equation}%
or%
\begin{equation}
\int\limits_{x_{1}}^{x_{2}}\left( y_{tt}\frac{dm}{dx}-C\sec ^{2}\theta 
\frac{d\theta }{dx}\right) dx=0  \label{main-alt}
\end{equation}%
is obtained. Again using $\frac{\partial ^{2}y}{\partial x^{2}}=\sec
^{2}\theta \frac{d\theta }{dx}$ and $\frac{dm}{dx}=\rho A_{0},$one gets%
\begin{equation}
\int\limits_{x_{1}}^{x_{2}}\left( \rho A_{0}y_{tt}-Cy_{xx}\right) dx=0
\end{equation}%
Since this must hold for all $x_{1},x_{2}$ we conclude that%
\begin{eqnarray}
\rho A_{0}y_{tt}-Cy_{xx} &=&0 \\
y_{xx} &=&\frac{\rho A_{0}}{C}y_{tt}
\end{eqnarray}%
which is the same as the previous result.

As a conclusion, we state that there is no need for any smallness
assumptions or first order approximations. Also notice that this equation is
valid for both finite and infinite length strings.

\section{On the Variability of Tension and Area}

Assuming now that the solution to the wave equation is somehow obtained, one
can get the solutions for the tension and cross-sectional area. The results
are%
\begin{equation}
T=C\sec \theta =C\sqrt{1+y_{x}^{2}}>0
\end{equation}%
and, from the conservation of mass equation,%
\begin{equation}
A(x)=A_{0}\frac{dx}{ds}=A_{0}\cos \theta =\frac{A_{0}}{\sqrt{1+y_{x}^{2}}}>0
\end{equation}%
It is interesting to note that the product $T(x,t)A(x,t)=CA_{0}$ is
conserved at all points and at all times. The tension and the
cross-sectional area at a point are inversely proportional to each other.
Another observation is that both depend on $y_{x}$ only.

Further, if $T$ is written as%
\begin{equation}
T=C\frac{ds}{dx}
\end{equation}%
then, one may argue that if at some time, $t^{\ast }$, a string has $ds=dx$
everywhere, then the tension would be the same everywhere, say $T_{0}$, and
one would necessarily have $T=C=T_{0}$. If such a configuration exists then
the general solution would have to be given as%
\begin{equation}
T=T_{0}\frac{ds}{dx}
\end{equation}

Note that the existence of a configuration in which $ds=dx$ at $t^{\ast }$
everywhere, implies $y_{x}(x,t^{\ast })=0$ everywhere. Then, the inescapable
conclusion is that $y(x,t^{\ast })$ is constant everywhere. Or, by a shift
of coordinates, one can argue that $y(x,t^{\ast })=0$, everywhere. Here, we
ignored the discussion of cases in which $ds=dx$ holds in finite intervals
of distinct displacements.

It can now be stated that such configurations can always be contrived to
exist: for example at those at times (say, $t<0$) at which the spring is
unloaded and at rest. If the spring is considered to be unloaded and at rest
prior to the application of initial conditions, then one may safely argue
that $y(x,t)=y_{t}(x,t)=0,$for all $x$, and $t<0,$ (simply consider energy
principles or the fact that this is a trivial solution of the string PDE,
and thus realizable). This also forces that $y_{x}(x,t)=0$ in the same
domain, giving $T(x,t)=T_{0}$ for all $x$, and $t<0$. As a result, $T_{0}$
is to be interpreted as the internal tension that would result had the
spring been unloaded and at rest. Note that this is not equal to the tension
in the initial condition.

\subsection{What Happens at $t=0$?}

\begin{figure}[htb!]
    \centering
      \includegraphics[width=0.8\textwidth]{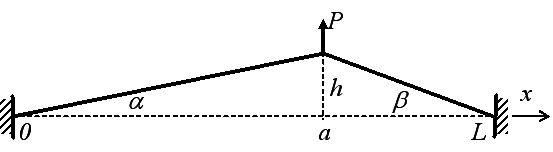}
    \caption{Initially displaced string under the action of a point force.}
  \label{Fig2}
\end{figure}

Another counter intuitive conclusion pertains to the situation at initial
condition. For any motion to ensue the string must be given an initial
displacement or velocity, or both. Let's consider a simple initial
displacement as shown in Figure \ref{Fig2}, with zero initial velocity.

The force $P$ is what is needed to induce the shown displacement. Based on
the results of this study, we can state that the tension at any point to the
left of the external force is $T_{L}=T_{0}\sec \alpha =T_{0}\sqrt{1+\left( 
\frac{h}{a}\right) ^{2}}$ and that to the right is $T_{R}=T_{0}\sec \beta
=T_{0}\sqrt{1+\left( \frac{h}{L-a}\right) ^{2}}$. Further, from the force
balance in vertical direction at point $a$ one gets%
\begin{eqnarray}
P &=&T_{L}\sin \alpha +T_{R}\sin \beta  \\
P &=&T_{0}\left( \tan \alpha +\tan \beta \right)  \\
P &=&T_{0}h\left( \frac{1}{a}+\frac{1}{L-a}\right) =\frac{L}{a\left(
L-a\right) }hT_{0}
\end{eqnarray}%
Now, imagine a quasi-static application of the vertical force starting from
zero, with zero deflection at point $a$, and gradually increasing to the
value of $P$, at which time the deflection at $a$ becomes $h$. In such a
process, one can simply integrate the work done by the vertical force in
order to determine the net energy stored in the string at the initial
condition. This gives%
\begin{equation}
U_{0}=\frac{1}{2}\frac{Lh^{2}}{a\left( L-a\right) }T_{0}=\frac{1}{2}Ph
\end{equation}

This is the internal energy stored that is converted into kinetic energy
when the motion ensues. One may even define an equivalent spring stiffness
at point $a$ as%
\begin{equation}
k_{\text{eff-}a}=\frac{L}{\left( L-a\right) }\frac{T_{0}}{a}
\end{equation}%
Thus, in response to vertical forces, current string model responds like a
linear spring, softest at the mid-point and stiffenning as $a$ aproaches to
the boundaries. Also, the stiffness is linearly proportional to the initial
tension. This behavior is quite familiar to those who have ever played a
plucked string instrument such as a guitar. Such results are impossible to
deduce from the classical treatments of ideal strings.

\subsection{Does It Make Sense?}

We immediately notice the situation of the string in regions where the slope
vanishes. In such portions the tension is simply equal to the tension at
rest. This could also be the case in initial condition. This is so even
though the tension at neighboring points just ouside such portions can
differ by a finite amount. For example, if there were two forces in the
previous figure that were equal and applied at $a=L/3$ and $b=2L/3$ then, as
the symmetry would require, the slope within middle section would have been
zero and the tension therein would have been equal to $T_{0}$.

This behavior is due to the assumption that the string points are allowed to
move only in transverse directions. The tension in the mid-section would
stay constant because there would be no extension in that section. In
reality, however, the tension in the mid-section would increase because some
material would leave the region at both ends due to higher tensions in the
first and third sections. Thus, in an analysis involving real material
behavior the points must be allowed to move in all directions.

\section{Energy Principles}

Previously, we determined the internal energy corresponding to an initial
displacement caused by a point force. In this section we present a
generalization that enables one to formulate the string problem using energy
methods. In addition, the results help validate the main results of this
study.

The internal energy is stored via changes in tension and length (area).
Since both of these depend only on spatial derivative of the displacement,
the energy stored while in motion will be the same as that in a static case,
as long as the displacement fields are identical. Therefore, given $%
u(x)=y(x,t=t^{\ast })$, where $t^{\ast }$ is a particular instant, one can
argue that the internal energy at this instant will be the same as the one
under the action of a suitable force distribution which, in static
equilibrium, induces the displacement field $u(x)$.

\begin{figure}[htb!]
    \centering
      \includegraphics[width=0.8\textwidth]{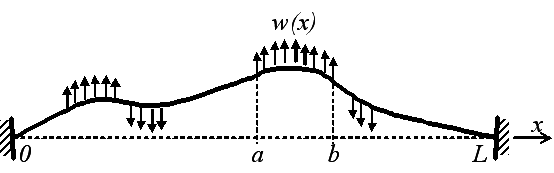}
    \caption{String displacement under the
action of a piece-wise continuous force distribution.}
  \label{Fig3}
\end{figure}

\begin{figure}[htb!]
    \centering
      \includegraphics[width=0.65\textwidth]{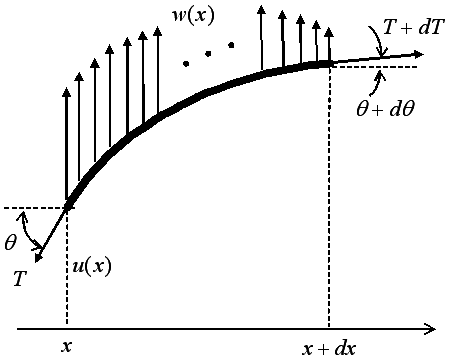}
    \caption{A differential string element in
transverse motion due to a continuous force distribution.}
  \label{Fig4}
\end{figure}

We now restrict the consideration to piece-wice smooth force distributions.
An example of this is shown in Figure \ref{Fig3}. For instance, $w(x)$ is
the external force distribution defined on the open interval $(a,b)$. Now,
imagine a small piece cut inside this domain, with internal tensions
appearing at cut ends, defined as before (see, Figure \ref{Fig4}). A
vertical force balance requires%
\begin{equation}
(T+dT)\sin \left( \theta +d\theta \right) -T\sin \theta =-\bar{w}(x)dx
\end{equation}%
where $\bar{w}(x)$ is a representative value, an average or that obtained by
applying the mean value theorem. Using $T=T_{0}\sec \theta $, this gives%
\begin{eqnarray}
\bar{w}(x) &=&-T_{0}\frac{\left( \tan \left( \theta +d\theta \right) -\tan
\theta \right) }{dx} \\
&=&-T_{0}\frac{\left( \tan \left( \theta +d\theta \right) -\tan \theta
\right) }{d\theta }\frac{d\theta }{dx}
\end{eqnarray}%
In the limit $\lim_{dx\rightarrow 0}\bar{w}(x)$ one obtains%
\begin{equation}
w(x)=-T_{0}\sec ^{2}\theta \frac{d\theta }{dx}
\end{equation}%
But, from Equation \ref{yxxsectheta}, this simply is%
\begin{equation}
w(x)=-T_{0}u_{xx}
\end{equation}

That is, the force distribution
necessary to induce a static deflection of $u(x)$ is proportional to $u_{xx}$%
. This also indicates that the smoothness of $u$ is required up to the
second derivative -- only in a piecewise manner.

Now, let $U$ be the internal energy of the string in configuration $u(x)$,
and $U_{0}$ be that in unloaded configuration at rest ($u(x)=0$,
everywhere). Then, by the first law of thermodynamics, we must require%
\begin{equation}
U-U_{0}=W(w)
\end{equation}%
where $W(w)$ is the work done by the external forces in bringing the string
from the initial to the final configuration.

Next, we aim at calculating this work done. For this, imagine that the
string is brought to the final configuration via a quasi-static process in
which all intermediate configurations are given by%
\begin{equation}
h(x)=\alpha u(x)\qquad \alpha \in \lbrack 0,1]
\end{equation}%
The corresponding force distributions would be given by: $w(x)=-\alpha
T_{0}u_{xx}$. The work done in going from $h$ to $h+dh$, due to a variation $%
d\alpha $, is given by%
\begin{equation}
dW=\left[ w(x)dx\right] dh=\left[ -\alpha T_{0}u_{xx}dx\right] (ud\alpha )
\end{equation}%
Thus, the net work done can be found from the integral%
\begin{equation}
W=\int_{a}^{b}\int_{0}^{1}\left( -T_{0}uu_{xx}\right) \alpha d\alpha dx=-%
\frac{1}{2}T_{0}\int_{a}^{b}uu_{xx}dx
\end{equation}%
where $a,b$ are the values of $x$ at boundaries, and either one can be taken
at infinity. The final integral can be simplified using integration by parts
as%
\begin{equation}
W=-\frac{1}{2}T_{0}\left[ \left. uu_{x}\right\vert
_{a}^{b}-\int_{a}^{b}u_{x}^{2}dx\right]
\end{equation}%
If one measures the internal energy by using $U_{0}$ as the reference, that
is $U-U_{0}\rightarrow U$, then the conclusion is%
\begin{equation}
U=\frac{1}{2}T_{0}\int_{a}^{b}u_{x}^{2}dx-\frac{1}{2}T_{0}\left( \left.
uu_{x}\right\vert _{a}^{b}\right)
\end{equation}%
This is the internal energy stored due to the work done by external forces.
When in motion, the same displacement field would represent the same
internal energy.

Now, we concentrate on the case of fixed boundaries (we mean, $u(a)=u(b)=0$)
for which the second term vanishes. For other cases refer to \cite{Weinstock}%
, where it is shown that the same equation of motion is obtained, only with
certain restrictions at boundaries. Hence, for fixed boundaries, we get%
\begin{equation*}
U(t)=\frac{1}{2}T_{0}\int_{a}^{b}y_{x}^{2}dx
\end{equation*}

Quite surprisingly, this is exactly the same as the one obtained in
classical treatments with constant tension assumption. For example, in \cite%
{Sagan} the internal energy is calculated using the assumption that the
string undergoes an extension under a constant tension, $\tau _{0}$, which
yields $U=\tau _{0}\int \left( \sqrt{1+y_{x}^{2}}-1\right) dx$. After
expanding the radical in series and ignoring the terms higher than the
linear, they get $U=\frac{1}{2}\tau _{0}\int y_{x}^{2}dx$. This is done even
after inextensibility is assumed! The question that remains is that whether
or not this is merely a lucky coincidence, despite so many conflicting model
assumptions in classical string models.

If, now, one writes $K(t)=\frac{1}{2}\int_{a}^{b}\rho ^{\ast }y_{t}^{2}dx$
for the kinetic energy, where $\rho ^{\ast }$ is the linear density, then
the Lagrangian of the system is%
\begin{equation}
\pounds (t)=K(t)-U(t)=\frac{1}{2}\int_{a}^{b}\left[ \rho ^{\ast
}y_{t}^{2}-T_{0}y_{x}^{2}\right] dx
\end{equation}%
which is the same as in \cite{Sagan}. Therefore, the same equation of motion
is obtained upon application of Hamilton's principle and variational methods
(see also \cite{Byron}, \cite{Weinstock}). This proves that the main results
of this study are also consistent with energy approaches.

\section{Variability of Area and Density at Rest}

Let us assume that for some reason the unloaded string has a spatially
varying cross-sectional area $A_{0}$ and density $\rho _{0}$. The
development presented using the alternative method is still valid up to
Equation \ref{main-alt}. Hence, we have%
\begin{equation}
\int\limits_{x_{1}}^{x_{2}}\left( y_{tt}\frac{dm}{dx}-Cy_{xx}\right) dx=0
\end{equation}%
Also, due to pure transverse motion assumption, the mass is conserved inside
any finite piece. Hence%
\begin{equation}
\int\limits_{x_{1}}^{x_{2}}dm=\int\limits_{x_{1}}^{x_{2}}\rho
(x,t)A(x,t)ds=\int\limits_{x_{1}}^{x_{2}}\rho _{0}(x)A_{0}(x)dx
\end{equation}%
or, arguing as before,%
\begin{equation}
\frac{dm}{dx}=\rho _{0}(x)A_{0}(x)
\end{equation}%
Thus,%
\begin{equation}
\int\limits_{x_{1}}^{x_{2}}\left( y_{tt}\rho
_{0}(x)A_{0}(x)-T_{0}y_{xx}\right) dx=0
\end{equation}%
and, one obtains%
\begin{equation}
y_{xx}=\frac{\rho _{0}(x)A_{0}(x)}{T_{0}}y_{tt}=\frac{\rho _{0}^{\ast }(x)}{%
T_{0}}y_{tt}
\end{equation}%
where $\rho _{0}^{\ast }$ is the linear density distribution at rest. This
is the most general equation of motion for transverse vibrations of a
string. Note again that in obtaining this result neither finiteness of
length nor any boundary conditions are argued. A similar result is obtained
in \cite{Weinstock}, although all other unnecessary assumption are still
used.

\section{Conclusion}

A new approach to the transverse motion of ideal strings is presented, in
which most of the approximations and assumptions of classical models are
removed. The new model allows variable tension and length, and, arbitrarily
large displacements and derivatives. Despite these relaxations the resulting
equation of motion is shown to be exactly the same as that of the classical
linear model. The result is obtained in three distinct ways: using
differential elements, finite elements, and energy principles. Another new
result relating force distributions to the second spatial derivative of
displacement is also presented. Finally, an equation of motion for variable
cross-sectional area and density is presented.

It must be cautioned here that, although the resulting equation of motion is
the same as that in previous studies, with the analyses presented in this
study we now have a variable tension and large deflection model.

\end{document}